\documentstyle[12pt]{article}

\renewcommand\baselinestretch{1.3}

\begin{document}

\begin{titlepage}
\renewcommand{\baselinestretch}{1}
\renewcommand{\thepage}{}
\title{\bf  The  thermohydrodynamical
picture  of a charged  Brownian particle. }
\author{L. A. Barreiro, J. R. Campanha  and R. E. Lagos \\
Departamento de F\'{\i}sica, IGCE UNESP\\
CP $178$,  $13500$-$970$ Rio Claro, SP, Brazil}

\date{}
\maketitle

\begin{abstract}
We study a charged Brownian gas with a non uniform bath
temperature, and present
a thermohydrodynamical picture. Expansion on the collision time probes
the validity of the
local equilibrium approach and the relevant thermodynamical variables. For
the linear regime we present several applications (including some novel results).
For the lowest nonlinear expansion and uniform bath temperature
we compute the gradient corrections to the local
equilibrium approach and the fundamental (Smoluchowsky) equation for the
nonequilibrium particle density.

\

\begin{flushleft}
 Nonequilibrium thermodynamics, hydrodynamical equations\\

\

{\em Corresponding author}: R. E. Lagos\\
Departamento  de  F\'{\i}sica- IGCE, Universidade Estadual Paulista (UNESP)\\
C.P. 178, Rio Claro 13500-970, SP, Brazil\\
Fax: (55)-19-534-8250,  email:monaco@rc.unesp.br
\end{flushleft}

\end{abstract}

\end{titlepage}

We generalize previous work \cite{bcl}, considering a magnetic field \cite
{garba} and a non uniform bath temperature $T({\bf x},t)$. Following \cite
{bcl} we write down Kramers equation for the probability distribution $P(%
{\bf x,v,}t)$ in phase space for the Brownian particle (mass $m$ charge $e$)
under an external force ${\bf F}$.

\begin{equation}
\left( \frac{\partial }{\partial t}+{\bf v}\frac{\partial }{\partial {\bf x}}%
+\frac{1}{m}{\bf F}\frac{\partial }{\partial {\bf v}}\right) P=J
\label{kkeq}
\end{equation}

\begin{equation}
J=\frac{1}{\tau }\frac{\partial }{\partial {\bf v}}\left( {\bf v}+v_{T}^{2}%
\frac{\partial }{\partial {\bf v}}\right) P-\frac{1}{\tau _{0}}\left( P-{\bf %
n}_{0}f_{0}({\bf v})\right)
\end{equation}

\noindent The Fokker-Planck collision time $\tau $ is the inverse of the
friction coefficient and the thermal velocity $v_{T}$ is given by $%
mv_{T}^{2}=T$ (hereinafter $k_{B}=1$). The diffusion coefficient $D$ and the
mobility $\xi =\tau /m$ are related via the celebrated Einstein relation $%
D=\xi T=\tau v_{T}^{2}$. In the collision term $J$ we include a generalized
BGK collision mechanism \cite{bgk}. The latter is characterized by a
relaxation time $\tau _{0}$ and a prescribed referential particle density $%
n_{0}({\bf x},t)$ (for example, as in \cite{bgk} $n_{0}$ is the actual
Brownian particle density $n({\bf x},t)$) and $f_{0}$ is the normalized
equilibrium distribution function

\[
f_{0}({\bf v})=\left( \frac{1}{2\pi v_{T}^{2}}\right) ^{\frac{3}{2}}\exp
\left( -\frac{{\bf v}^{2}}{2v_{T}^{2}}\right)
\]

The external force (not necessarily uniform) includes a mechanical part $-%
{\bf \nabla }V_{\mbox{mec}}$, an electric field ${\bf E=-\nabla
}\phi $ and
a magnetic field ${\bf B}$. Let us define a potential function $%
V=V_{mec}+e\phi $ and a covariant derivative ${\bf \nabla }_{cov}=$ ${\bf %
\nabla +\Gamma }$, where ${\bf \nabla }=\frac{\partial }{\partial {\bf x}}$
and ${\bf \Gamma }=T^{-1}{\bf \nabla }V$. Also define the vector ${\bf %
\omega }$ (notice the cyclotron frequency is $|{\bf \omega }|=\omega _{c}$)
by $mc{\bf \omega }=e{\bf B}$. Thus the external force ${\bf F}$ is given by
${\bf F(x,v)=-\nabla }V-m{\bf \omega \times v}$. Generalizing our previous
ansatz \cite{bcl} as

\begin{equation}
P=f_{0}({\bf v})\sum_{n_{i}=-\infty }^{\infty }\Psi _{{\bf n}}({\bf x}%
,t)\prod_{k=1}^{3}\left( \frac{\Phi _{n_{k}}(w_{k})}{\Phi _{0}(w_{k})}%
\right) ,\hspace{0.5cm}w_{k}=\frac{v_{k}}{\sqrt{2}v_{T}}  \label{ansatz}
\end{equation}

\noindent where $\Phi _{n_{k}}(w_{k})$ are the orthonormal Hermite functions
and with ${\bf n}=(n_{1},n_{2},n_{3})$. As before, we require the density of
Brownian particles to be $n{\bf (x,}t)=\int d{\bf v}P({\bf x,v},t)$. By
direct substitution of the ansatz into equation (\ref{kkeq}), and as in \cite
{bcl} a set of differential recursive (difference) equations for the $\Psi
_{n}^{\prime }s$ are readily obtained (in \cite{bcl}, the one dimensional
case, a continued fraction expansion was obtained, the general case is
amenable to a multibranched continued fraction scheme as in \cite{confrac}).
After a partial integration we obtain

\begin{equation}
\frac{\tau }{\tau _{0}}n_{0}\delta _{{\bf n,0}}+\tau {\bf \omega \cdot A}^{*}%
{\bf \times A}Z_{{\bf n}}=({\bf A}^{*}{\bf A+}\tau R)Z_{{\bf n}}
\label{recu}
\end{equation}

\begin{equation}
R=\partial +{\bf \nabla A^{*}+}v_{T}^{2}\left( {\bf \nabla }_{cov}{\bf A+}%
{\bf A}^{2}\left( \frac{\partial g}{\partial t}+{\bf Q\nabla }g\right)
\right)
\end{equation}

\noindent where

\begin{eqnarray*}
Z_{{\bf n}} &=&\frac{v_{T}^{n_{1}+n_{2}+n_{3}}\Psi _{{\bf n}}}{\sqrt{%
n_{1}!n_{2}!n_{3}!}} \\
\partial &=&\frac{\partial }{\partial t}+\frac{1}{\tau _{0}},\hspace{0.5cm}g=%
\frac{1}{2}\ln T
\end{eqnarray*}

\noindent The asymmetric lowering and raising operators ${\bf A}$, ${\bf A}%
^{*}$ and the displacement operator ${\bf Q}$ are defined respectively by

\begin{eqnarray*}
A_{1}Z_{{\bf n}} &=&Z_{n_{1}-1,n_{2},n_{3}} \\
A_{1}^{*}Z_{{\bf n}} &=&(1+n_{1})Z_{n_{1}+1,n_{2},n_{3}} \\
{\bf Q} &=&v_{T}^{2}{\bf A+A}^{*}.
\end{eqnarray*}

\noindent (similarly for the remaining components $A_{2}$ and $A_{3}$). The
particle number balance equation (equation (\ref{recu}) with ${\bf |n|}=0$)
defined here as the generalized Smoluchowsky equation, is given by:

\begin{equation}
\frac{\partial n}{\partial t}+{\bf \nabla J}=-\frac{1}{\tau _{0}}\left(
n-n_{0}\right)  \label{smolu}
\end{equation}

\noindent with ${\bf J=(}Z_{100},Z_{010},Z_{001}).$ The passage to
hydrodynamics is performed as before \cite{bcl} defining the particle flux,
the charge flux, the pressure tensor and the energy flux, respectively by:

\begin{eqnarray*}
{\bf J} &=&\int d{\bf vv}P({\bf x,v},t),\hspace{0.5cm}{\bf J}_{e}=e{\bf J} \\
{\bf \Pi (x,}t) &=&m\int d{\bf vvv}P({\bf x,v},t) \\
{\bf J}_{E}{\bf (x,}t) &=&\frac{1}{2}m\int d{\bf vvv}^{2}P({\bf x,v},t).
\end{eqnarray*}

We also define the energy density $E$ and the nonequilibrium temperature $%
\Theta $ as before \cite{bcl}

\[
E{\bf (x,}t)=\frac{1}{2}Tr{\bf \Pi =}\frac{3}{2}p=\frac{3}{2}n\Theta
\]

Again from equation (\ref{recu}), this time with ${\bf |n|}=1$, we derive
the balance equation for the charge flux (generalized Ohm law \cite{nic})

\[
\tau ^{*}\frac{\partial {\bf J}_{e}}{\partial t}+{\bf J}_{e}=\sigma
^{*}\left( {\bf E}+R_{H}{\bf J}_{e}\times {\bf B-}\frac{1}{e}{\bf \nabla }%
V_{mec}-\frac{1}{en}{\bf \nabla \Pi }\right)
\]

\noindent where

\[
\frac{1}{\tau ^{*}}=\frac{1}{\tau }+\frac{1}{\tau _{0}},\hspace{0.5cm}\sigma
^{*}=\frac{e^{2}\tau ^{*}n}{m},\hspace{0.5cm}R_{H}=\frac{1}{nec}
\]

\noindent With the additional definitions $\varepsilon =mnv_{T}^{2}$ and $%
\varepsilon _{0}=mn_{0}v_{T}^{2}$ we obtain for the energy flux balance
equation (equation (\ref{recu}), with $1<{\bf |n|}\leq 3$)

\[
\frac{\partial E}{\partial t}+{\bf \nabla J}_{E}=-{\bf J\nabla }V-\frac{2}{%
\tau }(E-\varepsilon )-\frac{1}{\tau _{0}}(E-\varepsilon _{0})
\]

We perform the passage to thermodynamics \cite{bcl} with the definition for
the entropy density

\[
S=\int d{\bf v}S^{*}({\bf x,v},t)=-\int d{\bf v}P({\bf x,v},t)\ln \lambda P(%
{\bf x,v},t),\hspace{0.5cm}\lambda =\frac{1}{e_{n}}\left( \frac{h}{m}\right)
^{3}
\]

\noindent (where $\ln e_{n}=1$, so in the appropriate limit we retrieve the
usual thermodynamical entropy \cite{bcl}). The entropy flux is defined as $%
{\bf J}_{S}=\int d{\bf vv}$ $S^{*}({\bf x,v},t)$, then the entropy density
balance equation is given by

\begin{equation}
\frac{\partial S}{\partial t}+{\bf \nabla J}_{S}=-\int d{\bf v}J\left( 1+\ln
\lambda P\right) =\sigma _{S}
\end{equation}

\noindent \noindent where $\sigma _{S\mbox{ }}$is the entropy
production Also as in \cite{bcl} we define the generalized free
energy densities $F$ and $G$, and the intrinsic and total
chemical potentials respectively as

\[
F=E-\Theta S,\hspace{0.5cm}G=F+p=n\mu _{int},\hspace{0.5cm}\mu =\mu _{int}+V
\]

\noindent Here we introduce no a priori assumptions on the functional
dependence of the entropy (as in the local equilibrium approach, LEQ) nor do
we assume relevant variables as is extended irreversible thermodynamics
(EIT, \cite{jou}). As in \cite{bcl}, all fluxes (including the ones not
computed here), $\Theta ,$ $\mu ,$ $S$, in fact all thermodynamic potentials
can be expanded in powers of $\tau $ (adequate at least in the overdamped
limit and with $\tau _{0}\gg \tau $), formally rendering all variables
universal functions of the set $\left\{ {\bf \nabla },{\bf D},n,T,{\bf B}%
\right\} $. Thus, at any given $\tau $ expansion stage, the only relevant
variables are $n$ and $T$. For Brownian motion, the universal equation for
the particle density is the generalized Smoluchowsky equation plus a set of
boundary conditions (BC). This picture is distinct but not inconsistent with
EIT \cite{jou}, where incorporating the BC to the fluxes, the latter may be
interpreted as relevant variables. LEQ is satisfied only to first order in $%
\tau $, where
\[
\Theta =T\left( 1-\tau \frac{\partial \ln g}{\partial t}\right) =T-\frac{%
\tau }{2}\frac{\partial T}{\partial t}
\]

For the linear (first order in $\tau $) case and a stationary (time
independent) temperature $T=\Theta $, the particle and heat fluxes are cast
as (with ${\bf B=}B\stackrel{\wedge }{z}$)

\begin{equation}
{\bf J=-K}_{1}\left( \xi n{\bf \nabla }V+{\bf \nabla }(Dn)\right)
\label{flux}
\end{equation}

\begin{equation}
{\bf J}_{E}=\frac{5}{2}T\left( {\bf J}-\xi n{\bf
K}_{\frac{1}{3}}{\bf \nabla }T\right)  \label{eflux}
\end{equation}

\begin{equation}
{\bf K}_{q}=\left(
\begin{array}{ccc}
\alpha _{q} & q\alpha _{q}\theta & 0 \\
-q\alpha _{q}\theta & \alpha _{q} & 0 \\
0 & 0 & q
\end{array}
\right) ,\hspace{0.3cm}\alpha _{q}=\frac{q}{1+(q\theta )^{2}}  \label{onsag}
\end{equation}

\noindent with $\theta =DD{_{B}}^{-1}$ and where Bohm's diffusion
coefficient is $D_{B}=cT{(eB)}^{-1}$. Notice that $\theta =\sigma
_{0}BR_{H}=\tau \omega _{c}$ where $\sigma _{0}=m^{-1}e^{2}\tau n$.

The scope of the Brownian motion scheme is highlighted with some direct
applications of equation (\ref{flux}):

a)If diffusion is disregarded (${\bf \nabla }n=0$, say electrons in a metal)
and with ${\bf \nabla }T=0$ we retrieve the classical magnetoresistance and
Hall effect models \cite{mermin};

b) If we consider two Brownian gases, say electrons and holes (inhomogeneous
semiconductors), with the $n_{0}$'s the equilibrium carriers' densities and
the $\tau _{0}$'s generation-recombination times, then (\ref{smolu})
together with (\ref{flux}) constitute generalized transport equations \cite
{mermin};

c)The diffusive part has the `correct Fick law' for nonuniform temperature
\cite{van};

d) Equation (\ref{flux}) can also de cast as

\begin{equation}
{\bf J}=-\xi {\bf K}_{1}(n{\bf \nabla }\mu +S{\bf \nabla }T)
\end{equation}

\noindent in agreement with \cite{landauer} for the ${\bf \nabla }T=0$ case,
suggesting that diffusive mass transport is solely convective transport of
both mass and entropy; and finally

e)In the high magnetic field regime and with ${\bf \nabla }T=0$ the cross
diffusion current is driven by D$_{B}\sim B^{-1}$ the anomalous Bohm
coefficient, never derived but proposed to explain the observed $B^{-1}$
behavior instead of the expected $B^{-2}$ term \cite{chen}.

f)Onsager's reciprocal relations are retrieved from equations (\ref{flux})
and (\ref{eflux}) only in the small magnetic field regime, with {\bf \ }$%
\alpha _{q}\approx q$ (see equation (\ref{onsag}).

Results up to $\tau ^{2}$ (lowest non linear regime) are presented only for
the uniform case ${\bf \nabla }T\equiv \frac{\partial T}{\partial T}\equiv
0. $ Define the magneto-covariant derivative ${\bf \vartheta =K}_{1}{\bf %
\nabla }_{cov}.$ Then the particle and energy fluxes are given respectively
by

\begin{eqnarray*}
{\bf J} &=&-D{\bf \vartheta }n-\frac{\tau D}{\tau _{0}}{\bf K}_{1}{\bf %
\vartheta }n_{0} \\
{\bf J}_{E} &=&\frac{5}{2}T{\bf J,}
\end{eqnarray*}

\noindent The nonequilibrium temperature is cast as

\[
\Theta =T\left( 1+\frac{\tau D}{3n}{\bf \vartheta }^{2}n\right)
\]

\noindent the nonequilibrium entropy, chemical potential and entropy
production are given respectively by

\[
S=S_{eq}(n,\Theta )-\frac{1}{2nv_{T}^{2}}{\bf J}^{2}=S_{eq}(n,\Theta )-\frac{%
\tau D}{2n}\left| {\bf \vartheta }n\right| ^{2}
\]

\[
\mu =\mu _{eq}(n,\Theta )+\frac{mD^{2}}{2n}\left| {\bf \vartheta }n\right|
^{2}
\]

\[
\sigma _{S}=\frac{D}{n}\left( \left| {\bf \vartheta }n\right| ^{2}-n{\bf %
\vartheta }^{2}n\right)
\]

\noindent (the latter for the case $\tau _{0}=\infty ).$ The universal
(Smoluchowsky) equation for the nonequilibrium density is

\[
\frac{\partial n}{\partial t}=D{\bf \nabla }\left( {\bf \vartheta }n+\frac{%
\tau }{\tau _{0}}{\bf K}_{1}{\bf \vartheta }n_{0}\right) -\frac{1}{\tau _{0}}%
(n-n_{0})
\]

Work in progress includes illuminated systems \cite{solar}, chemical
reactions \cite{tania}, the non linear regime in earnest, transport and
entropy production general properties, and the validity of any Onsager-like
scheme.

\noindent {\it Acknowledgments} This work was supported by {\em FAPESP, SP}
Brazil.


\begin{thebibliography}{99}
\bibitem{bcl}  L. A. Barreiro, J. R. Campanha and R. E. Lagos, arXiv:
cond-mat/9910405; Physica {\bf A 283 }(2000) 160.

\bibitem{garba}  R. Czopnik and P. Garbaczewsky, arXiv: cond-mat/0005353 \&
0011105.

\bibitem{bgk}  P. L. Bathnagar, E. P. Gross and M. Krook, Phys. Rev. {\bf 94}
(1954) 511.

\bibitem{confrac}  R. E. Lagos and R. A . Friesner, J. Chem. Phys.{\bf \ 81}
5899 (1984) 5899

\bibitem{nic}  D. R. Nicholson, Introduction to Plasma Theory (John Wiley \&
Sons, 1983)

\bibitem{jou}  D. Jou, J. Casas-V\'{a}squez and G. Lebon, Rep. Prog. Phys.
{\bf 62} (1999) 1035.

\bibitem{mermin}  N. W. Ashcroft and N. D. Mermin, Solid State Physics
(Holt, Rinehart and Winston, 1976).

\bibitem{van}  N. G. van Kampen, Z. Phys. {\bf B} 68 (1987) 135.

\bibitem{landauer}  R. Landauer, Helvetica Physica Acta {\bf 56} (1983) 847.

\bibitem{chen}  F. C. Chen, Introduction to Plasma Physics (Plenum Press, NY
\& London, 1974).

\bibitem{solar}  R. E. Lagos, H. Suhl and T. Tiedje, J. Appl. Phys. {\bf 54}
(1983) 3951.

\bibitem{tania}  R. E. Lagos, T. P. Sim\~{o}es and A. L. Godoy; Physica {\bf %
A} {\bf 257} (1998) 401.
\end{thebibliography}
\end{document}